\documentclass[prl,twocolumn,showpacs,amsmath,amssymb]{revtex4-1}

\usepackage{times}
\usepackage{graphicx}

\newcommand{\beq}{\begin{equation}}
\newcommand{\eeq}{\end{equation}}

\newcommand{\ba}{\begin{eqnarray}}
\newcommand{\ea}{\end{eqnarray}}

\newcommand{\jcap}{Journal of Cosmology and Astroparticle Physics}
\newcommand{\pasj}{Publications of the Astronomical Society of Japan}

\def\mnras{Mon. Not. R. Astron. Soc.}
\def\aap{Astron. Astroph.}
\def\apjl{Astroph. J. Lett.}

\def\gs{\mathrel{\lower0.6ex\hbox{$\buildrel {\textstyle >}\over{\scriptstyle \sim}$}}}
\def\ls{\mathrel{\lower0.6ex\hbox{$\buildrel {\textstyle <}\over{\scriptstyle \sim}$}}}

\begin{document}

\title{The speed of gravitational waves from strong lensed gravitational wave-electromagnetic signals }

\author{Xi-Long Fan$^{1,2}$}
\email{xilong.fan@glasgow.ac.uk}
\author{Kai Liao$^{3}$}
\author{Marek Biesiada$^{4,5}$}
\author{Aleksandra Pi{\'o}rkowska-Kurpas$^{4}$}
\author{Zong-Hong Zhu$^{1,5}$}
\email{zhuzh@bnu.edu.cn}
\affiliation{$^{1}$ School of Physics and Technology, Wuhan University, Wuhan 430072, China}
\affiliation{$^{2}$ Department of Physics and Mechanical \& Electrical Engineering, Hubei University of Education,  Wuhan 430205, China}
\affiliation{$^{3}$ School of Science, Wuhan University of Technology, Wuhan 430070, China}
\affiliation{$^{4}$ Department of Astrophysics and Cosmology, Institute of Physics, University of Silesia, Uniwersytecka 4, 40-007 Katowice, Poland}
\affiliation{$^{5}$ Department of Astronomy, Beijing Normal University, 100875, Beijing, China}


\begin{abstract}
We propose a new model-independent measurement strategy for the propagation speed of gravitational waves (GWs) based on strongly lensed GWs and their electromagnetic (EM) counterparts. This can be done in a two-fold way: by comparing arrival times of GWs and EM counterparts and by comparing the time delays between images seen in GWs and EM counterparts. The lensed GW-EM event is perhaps the best way to identify an EM counterpart. Conceptually this method does not rely on any specific theory of massive gravitons or modified gravity. Its differential setting (i.e. measuring the difference between time delays in GW and EM domains) - makes it robust against lens modeling details (photons and GWs travel in the same lensing potential)  and against internal time delays between GW and EM emission acts. It requires, however, that the theory of gravity is metric and predicts gravitational lensing similar as General Relativity.
We expect that such test will become possible in the era of third-generation gravitational-wave detectors, when about 10 lensed GW events would be observed each year. The power of this method is mainly limited by timing accuracy of the EM counterpart, which for kilonova is around $10^4$\,sec. This uncertainty can be suppressed  by a factor of $\sim 10^{10}$, if  strongly lensed transients of much shorter-duration associated with the GW event can be identified.  Candidates for such short transients include short gamma-ray burst and fast radio bursts.
\end{abstract}


\maketitle

\section{introduction}

Gravitational waves (GWs), which are the transverse waves of spatial strain, generated by time variations of the mass quadrupole moment of the  source and traveling at the speed of light, were predicted by Albert Einstein in \cite{1916SPAW.......688E}.   The first observational evidence for the existence of gravitational waves was made after the discovery of the binary pulsar system PSR 1913+16 by Hulse and Taylor \cite{1975ApJ...195L..51H} and its subsequent follow-up by Taylor and Weisberg \cite{1982ApJ...253..908T}.
Recent announcement of the first direct detection of gravitational waves (GW150914) by the advanced LIGO detector  \cite{2016PhRvL.116f1102A} was a great achievement which opened up a new window on the Universe. Moreover, the first GW signal detected in laboratory came from the merger of two massive black holes, proving the existence of these so far speculative binary systems. With GW detectors operating and gathering data one would also be able to test various aspects of gravitational physics, like the validity of General Relativity (GR), in a way unaccessible to other techniques.   For example, in alternative theories of gravity, the speed of GW could be different from the speed of light through the breaking of  weak equivalence principle or the existence of massive gravitons (see the review \cite{2014LRR....17....4W} and references therein). Indeed, the graviton Compton wavelength test has already been performed following the  first direct detection of GW \cite{2016arXiv160203841T} using the  dispersion measurement, as well as the Einstein’s equivalence principle test \cite{2016PhRvD..94b4061W,2016PhLB..756..265K}.

Binary neutron stars (NS-NS) are one of  promising sources to be routinely detected by the ground based detectors (such as advanced LIGO/VIRGO and the third generation detectors like the Einstein Telescope \cite{2012CQGra..29l4013S}. What makes them even more interesting is that they are expected to be accompanied by the electromagnetic (EM) counterparts  which could  be visible as kilonovae/mergernovae (see the review \cite{2015arXiv151205435F} and references therein) with peak r-band  magnitude $\sim22-25$ AB Mag (e.g. \cite{2015ApJ...814...25C}). They are  shorter-duration (of order of days) transient events similar to supernovae (SNe).   Short gamma-ray bursts (SGRBs), which have simple and sharp temporal features ( of order of  $0.1-1 s$), are another very promising EM counterparts of GW from NS-NS and NS-BH \cite[e.g.][]{2005Natur.437..845F}.  Because of the jet collimation $\sim10\%$ of the NS/NS systems will be aligned as to give an observable SGRB.
 Recently, the fast radio bursts  (FRB) have attracted considerable attention \cite{2013Sci...341...53T}.  The origin of FRB is not known, but they could also be the EM counterparts of GW from NS-NS and NS-BH \cite{2013PASJ...65L..12T}. These much shorter-duration transients (of order of $ms$) allows  us to reach the timing precision  $\sim 0.01 \;ms$ (e.g. FRB 130628 in \cite{2016MNRAS.tmpL..49C}). Detailed studies of the EM counterparts of GW signals focused on their properties, rates and identification strategies are the top research topics in current astrophysics from both theoretical and observational point of view.
Any improvement on the timing accuracy of the EM counterpart in the future will enhance the power of the method proposed in this paper.

Next generation GW detectors like the Einstein Telescope will improve an order of magnitude in sensitivity over the Advanced LIGO. This means that probed volume of the Universe will increase by three orders of magnitude. Perspectives for observing strongly lensed GWs from merging double compact objects (NS-NS, NS-BH, BH-BH) has been studied in \cite{2014JCAP...10..080B, 2015JCAP...12..006D,2016ApJ...821L..18P} with the prediction that Einstein Telescope should be able to detect several tens up to more than hundred of such events per year. This statistics is however dominated by BH-BH systems.  Although a pure BH-BH merger is not expected to has an EM counterpart, yet several  papers  \cite[e.g.][]{2016ApJ...827L..31Z,2016ApJ...819L..21L,2016PhRvD..94b4061W},  motivated by the plausible GBM transient associated with GW150914 \cite{2016ApJ...826L...6C}, have proposed the formation channels of EM  counterpart in BH-BH merger system and discussed their applications.

With new generation of dedicated surveys (e.g. SLACS, CASSOWARY, BELLS, SL2S), strong gravitational lensing has developed into a serious technique in extragalactic astronomy (galactic structure studies) and in cosmology. In this phenomenon, a source (typically a quasar or a distant galaxy) lensed by a foreground massive galaxy or cluster appears in multiple images. Light rays of these images travel along paths differing in length and probe gravitational potential of the lens at different depth experiencing different gravitational time delays. These two effects: geometrical and the Shapiro effect combine to produce the time delay between images \cite{1992grle.book.....S}.
If the source is intrinsically variable (and most quasars are), the light curves of its images can be used to extract out the time delay \cite{2013A&A...553A.120T}. This technique requires high-quality monitoring with sufficient cadence, season and campaign lengths so that the microlensing effects caused by the stars can be eliminated.  Moreover, quite recently the first  detection of the gravitationally lensed supernova  have been reported \cite{2015Sci...347.1123K} and the original dream of Sjur Resfdal came true \cite{1964MNRAS.128..307R}. In the case of lensed transient sources, like SNe, the measurements of time-delays between images can be much more accurate.
Besides its typical use to determine the Hubble constant \cite{2010ApJ...711..201S},
measurements of strong lensing time-delays has also been used to constrain the amplitude of the gravitational wave background \cite{1989PhRvL..63.2017A}.   The forthcoming Large Synoptic Survey Telescope (LSST) will find about $\sim8000$ lensed quasars, $\sim3000$ of which will be monitored and have the well-measured time delays in six frequency bands within ten years. The estimated number of robust time-delay measurements for these is around 400, each with precision $<3\%$ and accuracy $ 1\%$ \cite{2015ApJ...800...11L}. The LSST should also find some 130 lensed supernovae during its survey duration, while the deep, space-based supernova survey done by Joint Dark Energy Mission (JDEM) is expected to find $\sim 15$ lensed SN \cite{2010MNRAS.405.2579O}.  Note that, similar to SNe, the proposed isotropic counterparts to NS-NS mergers (e.g., kilonovae) have signals a few days of duration with a limited timing accuracy.

As already mentioned, one of the most important issues to be studied with GW detectors is testing the validity of GR, in particular the question whether GW travels with the speed of light. Similar questions arise within the Lorentz Invariance Violating (LIV) theories where the dispersion relation for the photon could be modified making the speed of light energy dependent.
The observed time of arrival delay between two events, such as emission of different energy photon-photon \cite{2016arXiv160201566W},   photon-neutrino \cite{1987PhRvD..36.3276L},  and GW-EM signals \cite{2004PhRvD..69j3502C,2003PhRvD..67b4015C}, have been proposed to constraint the respective propagation speed. The unknown intrinsic time delay in the emission time of such two signals to be compared contributes considerably to the uncertainty of the time of arrival method. Concerning speed of GW, it has been proposed in \cite{1998PhRvD..57.2061W} that using the phase information of the GWs from inspiralling compact binary  estimated by matched filtering technique, a bound on the graviton mass (hence on the speed of GW) could by made using GW alone. However, the expected bounds depend strongly on other physical effects relevant for the particular inspiralling system detected such as spin-induced precessions, orbital eccentricity, higher waveform harmonics, the merger-ringdown phase, etc.

Here, we propose a method to directly constrain the speed of GW  by using the strong lensing time-delays measured with GW and their EM counterparts. The differential setting of our method makes it free from the intrinsic time delays in the source (i.e. different emission times of GW and EM signal). General idea to use gravitationally lensed signals registered in GW and EM windows has independently been proposed by \cite{2016arXiv160205882C}. Our formulation is slightly different from theirs and is supported with more rigorous calculations. Also worth noting is the paper by \citet{2016arXiv160600458T} which claims that even within General Relativity it is possible for a lensed GW signal to come earlier than EM one (emitted simultaneously) due to wave effects in gravitational lensing (breakdown of geometric optics approximation). This result does not apply in our case where we consider galaxies acting as lenses.

\section{Method}

Our method is an extension of the idea proposed by \cite{2009MNRAS.396..946B} in the context of testing the Lorentz Invariance Violation by using energy dependence of time delays in gravitationally lensed systems. Let us assume that we observed a strongly lensed GW signal and identified its electromagnetic counterpart in the optical, radio waves or in gamma-rays. Then we would be able to measure time delays between the images independently in GW detectors -- $\Delta t_{GW}$ and in the electromagnetic window -- $\Delta t_{\gamma}$. They will be different if the speed of gravity $v_{GW}$ is different from $c$.
The difference $(\Delta t_{\gamma}- \Delta t_{GW})$ will bear information about the speed of GW. The bound on the $v_{GW}$ will have the following general form valid for a broad set of analytical lens models:
\begin{equation}\label{proof}
1 - \left( \frac{v_{GW}}{c} \right)^2 \! \leq \frac{\delta T}{\Delta t_{\gamma} F_{\rm lens}(z_l,z_s)},
\end{equation}
 where $\delta T$ is timing accuracy with which time delays are determined  and $F_{\rm lens}(z_l,z_s) \sim O(1)$ is some factor (weakly) dependent on the lens model and background cosmology (see below).   Let us stress that our method is purely empirical one: we do not assume any (non-existing so far as a consistent theory) model of massive gravitons. We just refer to differences in time delays in GW and EM windows, assuming, however, that the theory of gravity is metric and predicts gravitational lensing similar as General Relativity. It means we refer to purely classical, rather than quantum regime.
However, in order to be more specific in calculations we will assume below that gravitons are massive and travel along time-like geodesics.
In this sense the term ``graviton" should be perceived as a useful jargon rather than reference to the quantum nature of GWs.
There is a wide diversity of possible alternative theories of gravity not all of which will be well constrained by the method we propose. For some more recent reviews see, e.g. \cite{doi:10.1142/S0218271816410224} or \cite{2016arXiv160308955Y} where the constraining power of observed GW signals has been demonstrated and discussed.

{\it Propagation of massive gravitons on the cosmological background.}

The hypothesis that the speed $v_{GW}$ of GW could be different from $c$ means that gravitons should be treated as massive particles (having the rest mass $m_{GW}$) moving along timelike geodesics. Therefore their dispersion relation would be
\begin{equation}
E^2_{GW} - p_{GW}^2 c^2 = m_{GW}^2 c^4
\end{equation}
instead of
\begin{equation}
E^2_{\gamma} - p_{\gamma}^2 c^2 = 0.
\end{equation}
 as for the photons. Let us moreover assume that GW travel along radial geodesics in the flat Friedman-Robertson-Walker (FRW) model with the metric
 \begin{equation}
 ds^2 = c^2 dt^2 - a(t)^2 \left[ dr^2 + r^2 d\theta^2 + r^2 sin^2\theta d\phi^2 \right].
 \end{equation}
Generalization to non-flat FRW would be straightforward. Covariant and contravariant radial components of GW four momentum are related as:
\begin{equation}
p_r = a^2 p^r
\end{equation}
and obviously
\begin{equation}
\frac{dr}{dt} = \frac{p^r c^2}{E} = \frac{p_r c^2}{a^2 E}.
\end{equation}
Then it is easy to see that velocity of gravitons is
\begin{equation} v_{GW} = \frac{dr}{dt} = \frac{c}{a} \left[ 1 - \frac{1}{2} \frac{m_{GW}^2 c^2a^2}{p_r^2} \right].
\end{equation}
If the GW signal was emitted at the moment $t_e$ and detected (observed) at $t_0$, then the travel distance of GW is:
\begin{equation}
r_{GW} = r_{\gamma} - \Delta r_{GW},
\end{equation} where:
\begin{equation} \label{rgamma}
r_\gamma  =  \int_{t_e}^{t_0} \frac{c}{a(t)} dt = c \int_0^z \frac{dz}{H(z)},
\end{equation}  is the usual comoving distance to the GW source, and
\begin{equation} \label{Deltar}
\Delta r_{GW} = \frac{1}{2} \frac{m_{GW}^2 c^3}{p_r^2} \int_{t_e}^{t_0} a(t) dt .
\end{equation} Besides the expansion rate $H(z)$ we will also use its dimension less form $h(z)$ defined as $H(z) = H_0 h(z)$ where $H_0$ is the Hubble constant.
Using
\begin{equation}
p_r = a(t_e) \frac{E}{c},
\end{equation}
and  with the notation:
\begin{equation} I_n (z_1,z_2) := \int_{z_1}^{z_2} \frac{dz'}{(1+z')^nh(z')},
\end{equation}  one has:
\begin{equation} \label{deltar}
\Delta r_{GW} = \frac{1}{2} \frac{c}{H_0} \frac{m_{GW}^2 c^4}{E^2} (1+z)^2 I_2(0,z).
\end{equation}
The above formulae should be understood in the following way: if the emission time $t_e$ and detection time $t_0$ are fixed, i.e. the same for the GW and electromagnetic sources then the GW source is by $\Delta r_{GW}$ closer than electromagnetic source. On the other hand if they are emitting from the same location the GW signal would come by $\Delta t_{GW} = \Delta r_{GW} / c$ later than electromagnetic counterpart.
In other words, travel time for GW would be by $\Delta t_{GW}$ longer, as if the source was located by $\Delta r_{GW}$ farther.

{\it Strong lensing time delays}
For the purpose of illustrating our ideas we shall
restrict our attention to the singular isothermal sphere  (SIS) model which has been
proved to be a useful and reliable phenomenological model
of early type galaxies which dominate the population of lenses.
The generalization to
SIE (singular isothermal ellipsoids) and general power-law spherically symmetric
mass distribution
is rather straightforward and would not change our
conclusions.

The Einstein ring radius for the SIS model is:
\begin{equation} \label{Einstein SIS}
\vartheta_E =  4 \pi \frac{D_{ls}}{D_s} \frac{\sigma_v^2}{c^2}
\end{equation}
where $\sigma_v$ denotes one-dimensional velocity dispersion of
stars in lensing galaxy. If the lensing is strong i.e. the misalignment angle $\beta$ between the directions to the
lens and to the source is
$\beta < \vartheta_E$ then two co-linear
images A and B form on the opposite side of the lens, at radial
distances $\vartheta_A = \beta + \vartheta_E$ and $ \vartheta_B = \vartheta_E -
\beta$ having time delays between the images:
\begin{equation} \label{SIS delay}
\Delta t_{SIS} = \frac{1+z_l}{ 2 c} \frac{D_l D_s}{D_{ls}}(\vartheta_A^2 -
\vartheta_B^2)
\end{equation}
which according to the equations Eq.(~\ref{rgamma}) and Eq.(~\ref{Einstein SIS}) can
also be written as

\begin{equation} \label{SIS other way}
\Delta t_{SIS} = \frac{32 \pi^2}{H_0} \left( \frac{\sigma}{c} \right)^4 y
\frac{ \widetilde{r}(z_l)\widetilde{r}(z_l,z_s)}{\widetilde{r}(z_s)} ,
\end{equation}

where  $\tilde{r}(z_l)$ denotes  the dimensionless (i.e. with
$c/H_0$ factored out) comoving
distance to the lens and $y = \beta / \vartheta_E$.

In the context of massive photons it was shown by  \cite{1973PhRvD...8.2349L} that the bending angle is modified by a factor
$1+ \frac{m^2 c^4}{2 E^2}$. These considerations are valid in our case, so it means that impact parameters of photons and GW from the same image  are different and the Einstein angle gets modified to: $\vartheta_{E,GW} = \vartheta_E (1+ \frac{m_{GW}^2 c^4}{2 E^2})$. Therefore while calculating the time delay between images seen in GW, one has to consider this effect, which affects Shapiro time delay together with geometrical terms using corrections Eq.(\ref{Deltar}) in the distances $D_{ls}$ and $D_l$.

Now, we can see that the difference between image time delays observed in GW detectors and in the electromagnetic domain is
\begin{equation} \label{GWtimeDelay}
\Delta t_{SIS,GW} - \Delta t_{SIS,\gamma} = \Delta t_{SIS,\gamma}  
\frac{m_{GW}^2 c^4}{E^2}  F_{\rm lens}(z_l,z_s)
\end{equation}
where:
\begin{align}
F_{\rm lens}(z_l,z_s) &=
  1 +  \frac{(1+z_s) I_2(0,z_s)}{2 \widetilde{r}(z_l,z_s)}\nonumber\\
  &- \frac{(1+z_l) I_2(0,z_l)}{2 \widetilde{r}(z_l)} -  \frac{(1+z_l) I_2(0,z_l)}{\widetilde{r}(z_l,r_s)}
\end{align}

Therefore we could specify  the speed of GW through Eq. \ref{proof}   using the information from  the lensed GW-EM system  in term of a ``graviton'' :
\begin{equation}
\frac{m_{GW}^2 c^4}{E^2} = 1 - \left( \frac{v_{GW}}{c} \right)^2 ,
\end{equation}
If one would be able to measure such a difference in time delays this would also be a proof that gravitons are massive (i.e. that GR needs to be modified).

The accuracy $\delta T$ of time delay measurements
sets constraints on the  $v_{GW}$.   Assuming the galaxy-galaxy strong lensing system with $z_l=1$ and $z_s=2$ one has the following bound coming from the GW/EM difference in  lensing time delays
\begin{equation} \label{speed_bound}
1 - \left( \frac{v_{GW}}{c} \right)^2 \! \leq  4.26\; \times10^{-10} \!\left( \frac{\delta T}{1\;ms} \right)\! \left( \frac{\sigma}{250 \; km/s} \right)^{-4}\! \!\left( \frac{y}{0.1} \right)^{-1}
\end{equation}
where we also assumed $\Lambda$CDM cosmology with the Hubble constant $H_0 = 68 \; km\;s^{-1}\;Mpc^{-1}$, $\Omega_m =0.3$. Numerical factor setting the scale corresponds to lens velocity dispersion $\sigma = 250 \; km/s$, timing accuracy of $\delta T = 1\;ms$ and source - lens misalignment $y=0.1$
Recent discovery of ``Refsdal supernova'' \cite{2015Sci...347.1123K} and especially its reappearance \cite{2016ApJ...817...60T} demonstrated that we are starting discover transient events lensed by a cluster. The cluster-scale images have much bigger time delays than in the case of galaxy scale lenses. This means that in such case our method of differences in time delays would be more restrictive. For example, taking the value of time delay for the Refsdal supernova image SX which reappeared as predicted one would get a bound $ 1 - \left( \frac{v_{GW}}{c} \right)^2 \! \leq 3.2\; \times 10^{-11} $ assuming $1ms$ timing accuracy.

Apart from the limitation due to accuracy with which EM lensing time delay can be measured, the method described above is less restrictive than travel time techniques because it cannot take advantage of cumulative effect along the whole path. However, the strong lensing system seen both in EM and GW offers additional possibility to compare the moments of arrival of the same image  seen in the EM and GW respectively. This would be possible only for transient EM sources like kilonova or better yet, SGRB associated with GW signal.  Then, according to the Eq.(\ref{Deltar}), the expected time delay (in each image) would be:

\begin{equation}
\Delta t_{\gamma,GW} = \frac{1}{2 H_0} (1+z_s)^2 I_2(0,z_s)
\end{equation}
For the source at the redshift $z_s = 2$ one has the following bound: $ 1 - \left( \frac{v_{GW}}{c} \right)^2 \! \leq 9.92\; \times 10^{-22} $

It would be appropriate to compare the above bounds with the results published in \cite{2016PhRvL.116f1102A} and \cite{2016arXiv160203841T} concerning the constraints on violations of general relativity leading to massive gravitons. The bound obtained from the GW150914 event was formulated in terms of graviton Compton wavelength $\lambda_{GW} > 10^{13} \;km$ which turned out to be the strongest \textit{dynamical} bound probing the propagation of gravitational interactions. Translating this into a bound on the speed of gravity, one obtains: $ 1 - \left( \frac{v_{GW}}{c} \right)^2 \! <  10^{-19}$. This constraint is much stronger than one can get from differences in time delays. Let us remind, however, that the aforementioned bound was obtained as a result of sophisticated analysis using waveform models that allow for parameterized general-relativity violations during the inspiral and merger phases and using the dispersion measurement.
On the other hand, the second method discussed by us - using the GW vs. EM arrival times in lensed images - is by three orders of magnitude more restrictive. This means that even a single instant of gravitationally lensed GW signal accompanied by EM transient counterpart would be valuable.


\section{Perspectives}

One can expect that the next decades of observations carried together in the GW and EM windows will be sufficient to give a strong constraint on the GW speed and graviton mass. Concerning our method, its main limitation is the accuracy of EM time delay,  while timing in the GW detectors is very precise ($< 10^{-4}\; ms$).

The planed third generation gravitational wave detector, such as the Einstein Telescope, could observe the strongly lensed GW. The rate of yearly detections of strongly lensed GW from NS-NS and NS-BH sources are in the range $\sim$2 to 10/yr \cite{2014JCAP...10..080B,2015JCAP...12..006D}, depending on different ET  configurations, stellar population synthesis models \cite{2013ApJ...779...72D}.  Cadenced wide-field EM imaging surveys in the next decade will increase the catalog of strongly lensed systems by two orders of magnitude. Besides, one can imagine a dedicated follow-up project based on the observed GW events.
Short-duration EM counterparts transients (such as kilonova) have a strong and pronounced feature on the light curve ( i.e. the maximum point), which creates a unique opportunity for time-delay extracting algorithms, resulting in a  accurate estimate. For these objects, we expect to   get  the time delay precision $\sim 10^{4}\; s$  through the dedicated photometry related to maximum point.   Much shorter-duration EM counterparts,  like SGRBs, FRBs or any new signal  discovered in the future, will be measured with  much better  time delay accuracy.   The constraint on GW speed can be enhanced by increasing the number $N$ of lensed systems observed in GW and EM windows. In such a case, the statistical uncertainty  would be reduced by a factor of $\sqrt{N}$.
Such a  population of  much shorter-duration EM counterparts, such as  $\sim$ 10 FRBs with $0.01ms$ time delay accuracy,   could suppress  the uncertainty of time delays   by a factor of  $\sim10^{10}$ comparing with  measured by a kilonova. The issue of expected rates of joint EM/GW strongly lensed events is interesting on its own and merits further studies. However, even a single such event -- discovered either serendipitously or as a result of dedicated surveys -- would be very important.

Our approach has a number of advantages. First is its differential setting making it robust, as already mentioned. However, the price paid for this is that it is much less restrictive. Second, the lensed EM/GW event is perhaps the best way to identify an EM counterpart: even with poor resolution of GW detectors if we see a lensed GW (two strains of similar temporal structure) coincident with lensed EM source we can be almost sure about the source location.  Extra bonus, then is that besides differential time delays we would be able to measure time of flight differences GW vs. EM in each image. If there are more than two images - e.g quads which are typical in strong lensing systems discovered so far, we could have several measurements from a single lensed source.
One has to remark, however, that as discussed in \cite{2015ApJ...798L..36C}
the peak amplitude of GW emission associated with the time of the merger could be registered long before the EM prompt SGRB signal. The two signals would be separated by the lifetime of the supramassive  NS,  which  can  easily  exceed $10^3\;s$. The intrinsic time delay between EM and GW signal is very hard to disentangle from the possible time delay due to hypothetical difference between the speed of light and speed of gravity. Therefore it is a serious obstacle to the method of EM/GW time of flight differences in each image. 

We can conclude that according to the anticipated development of GW astrophysics, massive EM surveys and the synergy between them will create possibility to use strongly lensed GW-EM events as complementary tests of fundamental physics and astrophysics.When this Letter was under review we became aware of the paper \cite{2016arXiv161202004B}, in which the authors independently discussed the idea and applications of multi-messenger time delays from lensed GWs.

\section{acknowledgments}

The authors would like to thank the referees for their valuable comments which allowed to improve considerably the original text.
X. F. thanks M. Sereno and Y. Chen for  valuable comments.
X. F. was supported by the National Natural Science
Foundation of China (No. 11303009 and 11673008) and Newton International Fellowship Alumni follow on funding. K. L. was supported by the NSFC No.  
11603015. M. B. obtained approval of
foreign talent introducing project in China and gained special fund
support of foreign knowledge introducing project.
Z. Z. was supported by the NSFC  No. 11633001.
This research was also partly supported by the Poland-China Scientific \& Technological
Cooperation Committee Project No. 35-4. 
\bibliography{bibliography_new_re.bib}


\end{document}